\documentclass[twocolumn,preprintnumbers,prb,aps,a4paper,amsmath,amssymb]{revtex4}

\newlength\picturewidth
\usepackage[latin1]{inputenc}
\usepackage{amsmath}
\usepackage{amssymb}
\usepackage{amscd}
\usepackage{graphicx}
\usepackage{subfigure}
\usepackage{epstopdf}
\usepackage{graphicx}% Include figure files

\def\reefree{R_\text{F}}
\def\rg{R_{\text{g}}}
\begin{document}

%%%%%%%%%%%%%%%%%%%%%%%%%%%%%%%%%%%%%%%%%%%%%%%%%%%%%%%%%%%%%%%%%%%%%%%%%%%%%%
%                    Title Page
%%%%%%%%%%%%%%%%%%%%%%%%%%%%%%%%%%%%%%%%%%%%%%%%%%%%%%%%%%%%%%%%%%%%%%%%%%%%%%
\title{Unexpected relaxation dynamics of a self-avoiding polymer in cylindrical confinement}

\author{Axel Arnold}
\email{arnold@amolf.nl}
\affiliation{FOM-Institute AMOLF, Kruislaan 407, 1098 SJ Amsterdam, The Netherlands}

\author{Behnaz Bozorgui}
\affiliation{FOM-Institute AMOLF, Kruislaan 407, 1098 SJ Amsterdam, The Netherlands}

\author{Daan Frenkel}
\affiliation{FOM-Institute AMOLF, Kruislaan 407, 1098 SJ Amsterdam, The Netherlands}

\author{Bae-Yeun Ha}
\affiliation{Department of Physics and Astronomy, University of Waterloo, Ontario N2L 3G1, Canada}

\author{Suckjoon Jun}
\affiliation{FAS Center for Systems Biology, Harvard University, 7 Divinity
  Avenue, Cambridge, MA 02138, USA}

\date{\today}

\begin{abstract}
  We report extensive simulations of the relaxation dynamics of a self--avoiding
  polymer confined inside a cylindrical pore.  In particular, we concentrate on
  examining how confinement influences the scaling behavior of the global
  relaxation time of the chain, $\tau$, with the chain length $N$ and pore
  diameter $D$.  An earlier scaling analysis based on the de Gennes blob picture
  led to $\tau \sim N^2 D^{1/3}$. Our numerical effort that combines molecular
  dynamics and Monte Carlo simulations, however, consistently produces different
  $\tau$--results for $N$ up to $2000$. We argue that the previous scaling
  prediction is only asymptotically valid in the limit $N \gg D^{5/3} \gg 1$,
  which is currently inaccessible to computer simulations and, more
  interestingly, is also difficult to reach in experiments.  Our results are
  thus relevant for the interpretation of recent experiments with DNA in nano--
  and micro--channels.
\end{abstract}
\preprint{relaxation in cylindrical confinement}
\maketitle

%%%%%%%%%%%%%%%%%%%%%%%%%%%%%%%%%%%%%%%%%%%%%%%%%%%%%%%%%%%%%%%%%%%%%%%%%%%%%%
%                 Begin main text
%%%%%%%%%%%%%%%%%%%%%%%%%%%%%%%%%%%%%%%%%%%%%%%%%%%%%%%%%%%%%%%%%%%%%%%%%%%%%%

\section{Introduction}

Polymer chains immersed in solution are subject to constant molecular collisions
and restlessly undergo conformational changes.  Since the motion of each monomer
or chain segment on a polymer is influenced by the rest, the polymer shows
unique dynamical properties compared to those of simple molecules. Confinement
in a pore or slit can change both the static and dynamic properties of a polymer
chain qualitatively~\cite{BdG}. In the presence of cylindrical confinement, for
instance, the pore diameter enters as an additional length scale. An early
scaling approach~\cite{BdG} shows how this alters the relaxation dynamics.

One of the key concepts concerning polymer dynamics is the rate at which the
modes relax~\cite{DE,deGennesBook,Rubinstein,Grosberg}.  Crudely speaking, this
is a measure of the relative importance of each mode in describing the {\em
  internal motion} of a chain molecule.  In particular, the longest relaxation
time or the global relaxation time $\tau$ is of practical importance, as it
determines how fast the chain reaches its equilibrium. Beyond this time, the
internal motion plays no significant role, and the chain behaves just like a
simple molecule.

A simple model for a polymer is the ``ideal'' chain, which is a phantom chain
without self-avoidance. The Rouse model provides a description of its relaxation
dynamics in an ``immobile'' solvent without hydrodynamic effects, and has been
extended to the case of a free self--avoiding chain in the presence or absence
of hydrodynamic effects~\cite{DE,deGennesBook,Rubinstein,Grosberg}. In a similar
spirit, the scaling behavior of a self--avoiding chain trapped in a pore of
diameter $D$ has been studied~\cite{BdG}; the trapped chain is viewed as a
linear string of \emph{compression blobs}~\cite{deGennesBook,BdG,Rubinstein} of
diameter $D$.
          
The problem of a self-avoiding polymer under cylindrical confinement has
relevance in a variety of contexts: DNA manipulations in nano-- or
micro--channels~\cite{reisner}, bacterial chromosome
segregation~\cite{future,Jun06}, and polymer translocation through narrow
pores~\cite{Kasianowicz02, Dekker}, to name a few. Additionally, the motion of
such a polymer is reminiscent of reptation in concentrated polymer
solutions~\cite{deGennesBook,DE}. Consequently, this problem has been
investigated in a number of simulation
studies~\cite{Takano,Jendrejack,chakraborty,sheng,KB} before.  However, to our
knowledge, the asymptotic blob scaling prediction~\cite{BdG} for $\tau$ has
never been \emph{directly} confirmed.

In this work, we perform extensive simulations of a cylindrically confined
chain, focusing on the slowest relaxation time $\tau$.  The main goal is to
examine dependence of $\tau$ on chain length $N$ and pore diameter $D$. For
simplicity, we ignore hydrodynamic effects.  Even with this simplification, the
analysis of $\tau$ is quite nontrivial, as detailed below.  In an effort to
present a concrete picture, we combine molecular dynamics (MD) and Monte Carlo
(MC) simulations.  While the MD simulations permit a more direct probe into
polymer dynamics, the MC simulations allow us to consider longer chains.

Despite the fact that we investigate a wide parameter space, we observe only an
``intermediate'' regime in our simulations, in which $\tau$ can be described by
power laws with a much stronger $D$--dependence and a weaker $N$--dependence
than the blob--scaling prediction. Interestingly, finite-size effects are more
sensitively reflected in the dynamical quantity $\tau$; the static intermediate
regime is much narrower than the dynamic counterpart. This intermediate regime
might be relevant for the interpretation of single--molecule experiments on
confined polymers, e.~g., DNA in a nano-- or micro--channel~\cite{reisner}.

This article is organized as follows. In Sec.~\ref{sec:theory}, we present a
simple physical picture for describing the scaling behavior of the global
relaxation time $\tau$ of a chain trapped in a cylindrical pore.  The simulation
methods employed in this work are described in Sec.~\ref{sec:sim}.
Section~\ref{sec:eeresults} is devoted to a detailed discussion on the results
of the MD simulations and their analysis.  In Sections \ref{sec:historesults}
and \ref{sec:blobresults}, we present results of our lattice (MC) simulations of
the end-to-end distribution of the confined chain, from which we extract $\tau$
and compare it with the MD result presented in earlier sections. Finally, we
discuss in Section~\ref{sec:discussion} the implications of our results for
single--molecule experiments on confined polymers.

\section{Theoretical background: Blob model and beyond}
\label{sec:theory}

\begin{figure}[tp]
  \centering
  \includegraphics[width=\picturewidth]{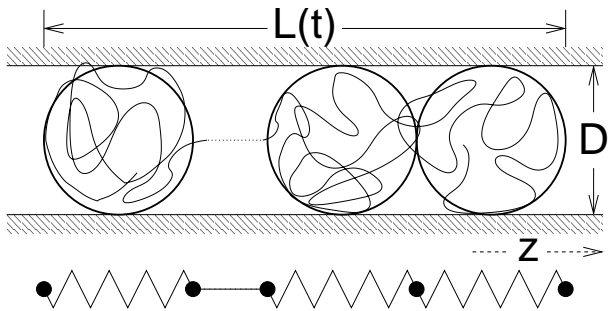}
  \caption{
    Cylindrically confined self--avoiding chain.  In the blob picture, the chain
    is viewed as a linear string of compression blobs depicted as spheres.
    Inside each blob of diameter $D$, the effect of wall confinement is
    considered to be minor; beyond $D$, it tends to align the chain in the
    longitudinal direction, i.~e., along the $z$ axis in our convention.  For
    small deformations, the blobs are assumed to be deformed independently of
    each other, as illustrated by the series of springs in the lower figure. }
  \label{fig:spring}
\end{figure}

Consider a linear self-avoiding chain consisting of $N$ monomers of size $a$,
trapped in a cylindrical pore of diameter $D$ (see Fig.~\ref{fig:spring}). In
our convention, the $z$--axis coincides with the symmetry axis of the cylinder.
If the longitudinal chain extension $L$ is much larger than $D$ and $D\gg a$, we
can use the de Gennes blob model, which views the elongated chain as a string of
self--avoiding compression blobs of diameter
$D$~\cite{deGennesBook,BdG,deGennes}.

Within a blob, the chain is expected to resemble a free or unconfined
self-avoiding chain of $g$ beads.  By equating $D$ with the ``Flory radius'' of
the blob~\cite{deGennesBook,DE,Flory} that scales as $a g^\nu$, one obtains $g
\sim (D/a)^{1/\nu}$, where~\cite{footflory} $\nu\approx 0.59$.  Linear (i.~e.,
longitudinal) ordering beyond $D$ demands that the total equilibrium or average
chain extension $L_\text{eq}$ should scale as
\begin{equation}
  \label{eq:chainlen}
  L_\text{eq} = \left<L \right> \sim (N/g)D \sim N a^{\frac{1}{\nu}} D^{1-\frac{1}{\nu}},
\end{equation}
where $\left< ... \right>$ denotes an ensemble average. This scaling behavior
has recently been confirmed in a theoretical approach~\cite{thirumalai}.

To estimate the longest relaxation time $\tau$ of the chain, we assume a
Hookian--spring--like response to small excitations from the equilibrium length
$L_\text{eq}$. If we know the effective spring constant $k_\text{eff}$ of the
chain, the relaxation time is then reciprocally obtained as $\tau \sim
N/k_\text{eff}$, where we have assumed that the chain friction is additive.  For
such small excitations, the spring constant of each blob should
be~\cite{BdG,blob_spring_const} $k_b \simeq k_BT/D^2$; here and in what follows,
$k_B$ is the Boltzmann constant and $T$ is the temperature.  We thus obtain
\begin{equation}
  \label{eq:keff}
  k_\text{eff} \sim \frac{k_b}{N_\text{blobs}} \sim
  \frac{k_BT}{a^2 N \big(\frac{D}{a}\big)^{2-\frac{1}{\nu}}},
\end{equation}
where $N_\text{blobs}=L_\text{eq}/D=N/g$ is the total number of blobs in the chain.  

The mapping of the confined chain onto a series of Hookian springs used in
Eq.~\ref{eq:keff} can be understood in analogy with a Rouse chain -- beyond $D$,
the confined chain resembles a one-dimensional Rouse chain~\cite{BdG,Takano}.
Slow Rouse modes are analogous to a random walk in a harmonic potential whose
spring constant varies as $1/N$.  The only difference is that in the confined
case $N_\text{blobs}$ should be used in place of $N$, because now the blob--size
$D$ is the smallest length scale for one-dimensional ``Rouse deformations''.

Then, the global relaxation time is
\begin{equation}
  \label{eq:trelax}
  \tau \sim \frac{\zeta N}{k_\text{eff}} \sim
  \frac{\zeta a^{\frac{1}{\nu}}}{k_BT} N^2 D^{2-\frac{1}{\nu}},
\end{equation}
where $\zeta$ is again the friction constant of each monomer. This derivation
recovers the well-known result of Brochard and de Gennes~\cite{BdG} based on the
blob picture.  Hereafter, for simplicity, we use the following conventions:
$\zeta = 1, a=1, k_BT = 1$. Then Eq.~\eqref{eq:chainlen} becomes $
L_\text{eq}=ND^{1-1/\nu}\approx ND^{-0.7}$, while Eq.~\eqref{eq:trelax} reduces
to~\cite{footexp} $\tau \sim N^2 D^{2-\frac{1}{\nu}}\approx N^2 D^{0.3}$.

The effective spring constant $k_\text{eff}$ of a chain (or any elastic rod) can
be obtained from its force-extension relation, from which the relaxation time
$\tau$ can be deduced (Eq.~\ref{eq:trelax}).  Alternatively, one can relate
$k_\text{eff}$ to the equilibrium distribution of the end-to-end distance $L$,
denoted by $p(L)$. The free energy cost for a small change in the chain length
is a long--time potential effectively felt by the chain and determines the
equilibrium distribution $p(L)$, which enables us to relate $k_\text{eff}$ to
$p(L)$.  For $L \approx L_\text{eq} = \left< L \right>$, the distribution $p(L)$
is Gaussian
\begin{equation}
  \label{eq:gauss}
  p(L) = \frac{1}{\sqrt{2 \pi} \sigma_L} \exp
  \left[- \frac{(L - L_\text{eq} )^2}{2 \sigma_L^2}\right],
\end{equation}
where $\sigma_L$ is the variance of the distribution.  This naturally arises
from the assumption of a Hookian--spring--like response; the free energy should
be $\sim (L-L_\text{eq})^2$ for $L\approx L_\text{eq}$.  For large deformations,
i.~e., $L \ll L_\text{eq}$, however, the distribution does not have to remain
symmetrical with respect to the line $L = L_\text{eq}$, as assumed in
Eq.~\ref{eq:gauss}; see Sec.~\ref{sec:historesults} for details.  The free
energy cost for a global deformation is then $-k_BT \ln p(L)$, and one
can thus establish $k_\text{eff} \simeq \left( \partial^2/\partial L^2 \right)
\left[ k_BT \ln p(L) \right] = \sigma_L^{-2}$, which agrees with the
equipartition theorem~\cite{Landau}.  The relaxation time $\tau$ can then be
related to the static quantity $\sigma_L$ as
\begin{equation}
  \label{eq:tauvariance}
  \tau \sim \frac{N}{k_\text{eff}} \sim N \sigma_L^2.
\end{equation} 
Thus, the computation of $\tau$ boils down to that of $\sigma_L$.  

\section{Simulation techniques}
\label{sec:sim}

For the direct, dynamical measurements of the relaxation times, we employed an
off-lattice Molecular dynamics (MD) simulation scheme together with a Langevin
thermostat for a bead-spring model; to determine static properties such as the
end-to-end distance distribution, we performed a lattice Monte-Carlo (MC)
simulation that allows for efficient sampling of chain conformations.  Below, we
describe both simulation techniques.

\subsection{MD details}

In the MD simulations, we used a bead-spring model of polymers, which were
trapped inside a long cylindrical tube with a circular or square cross section
(we refer to the latter as ``brick-shaped'' confinement).  The bead-bead and
bead-wall interactions were modeled by the Weeks-Chandler-Andersen (WCA)
potential~\cite{wca} (i.e., the repulsive part of the Lennard-Jones potential):
\begin{equation}
  U_\text{WCA}(r)=\epsilon_\text{WCA}\left[\left(\frac{a}{r}\right)^{12} -
    \left(\frac{a}{r}\right)^6 + \frac{1}{4}\right]
  \label{eq:WCA}
\end{equation}
for $r<\sqrt[6]{2}a$ and 0 otherwise.  Here $r$ denotes the center-to-center
distance between two beads, or the distance of a bead center from the confining
cylinder minus $a$ (i.e., the cylinder wall excludes the centers-of-mass of the
beads). With our choice of $\epsilon_\text{WCA}=1\; k_BT$, this potential models
soft beads of diameter $a$: their centers cannot come much closer to each other
than $a$ and are confined to be on the inside of the confining wall. In the
simulation, $a$ defines the basic length scale and $\epsilon_\text{WCA}$ the
energy scale, i.e., we choose $a=1$ and $\epsilon_\text{WCA}=1 k_BT$; in
addition, we fix the mass scale so that the mass of each bead $m=1$.  This
automatically sets our basic time scale as $\tau_\text{WCA}=a
\sqrt{m/\epsilon_\text{WCA}}=1$. Following the conventions in the theory
section, we will henceforth omit the units of these quantities.

The bond between two neighboring beads, which endures chain connectivity, was
modeled by the FENE (finite extensible nonlinear elastic) potential~\cite{FENE}
\begin{equation}
  U_F(r) = -\frac{1}{2} 4 \epsilon_F \ln \left[ 1 - \left(
      \frac{r}{2} \right)^2 \right]\,,
  \label{eq:fene}
\end{equation}
where $r$ is again the distance of the bead centers and $\epsilon_F$ is the
interaction strength. In the present simulations, we chose $\epsilon_F=10$,
which in combination with the our WCA potential results in a typical bond length
of $1.027$.

We simulated this system using the simulation package ESPResSo~\cite{espresso}.
To integrate the equations of motion, we employed a velocity-Verlet MD
integrator with a fixed time step of $0.01$; the system was kept at a constant
temperature by means of a Langevin thermostat with a fixed friction of
$\zeta=m\tau_\text{WCA}^{-1}=1$. The effect of the thermostat is {\em not only}
to keep the temperature constant {\em but also} to ensure that each monomer
moves diffusively, rather than ballistic. This is to mimic the effects of
solvent viscosity in the high-damping limit, in which the inertia term can be
dropped~\cite{DE}.

Initially, the chain was created as a biased, non-self-avoiding random walk
confined within the pore.  To equilibrate the system and remove possible
bead-bead overlaps, we simulated the system for a few thousand steps with a
truncated WCA potential. In other words, the WCA potential was modified such
that the potential is linear for distances smaller than a certain cutoff radius.
We reduced this cutoff gradually, until the potential had eventually converged
to the full WCA interaction.

Next, we equilibrated the system further for a total equilibration time of
$T_{\text{eq}}$, until the end-to-end distance and the radius of gyration had
converged to their steady averages. This was followed by the sampling phase with
a duration of $T_{\text{total}}$, during which we periodically recorded
configurations for analysis. Between each two recorded configurations we waited
for a time of $T_{\text{sample}}$ to reduce the statistical dependence of the
configurations. For the parameters of the performed simulations, see
Appendix~\ref{sec:simparams}.

\subsection{MC details}

In an effort to sample a wide parameter space, we complement our MD simulations
with lattice-based MC simulations.  In our MC simulations, the polymers were
modeled as a self-avoiding walk on a cubic lattice between hard walls forming
the cylinder.  In contrast to the MD simulations, where we typically used a
cylindrical pore with a circular cross section, we mostly used one with a square
cross section of area $D^2$ in the MC simulations. The latter is more compatible
with the geometry of the cubic lattice (on which the polymer lives).
Additionally, hard-core repulsions were defined as excluding conformations where
two monomers occupy the same lattice site.

For the Monte Carlo moves, we adopted the ``wormhole'' method of
Houdayer~\cite{wormhole}. This algorithm consists of a reptation-like motion so
that the first bead jumps through a ``wormhole'' to a random position --- not to
be confused with reptation as a physical process in concentrated
solutions~\cite{DE,deGennesBook}.  During the wormhole move, the polymer
therefore consists of two disconnected segments. The reptation is continued,
until a valid (i.e., connected) polymer conformation is obtained. This algorithm
is known to be very efficient for sampling the chain conformations of polymers
with excluded volume in confinement or in dense polymer melts. We only use this
wormhole method to study static properties, since, by its nature, the dynamics
based on this method is completely unphysical.

Between two chain conformations, we performed 1000 wormhole steps to assure
statistical independence, and sampled in total 20,000 conformations per
parameter set. Note that this algorithm requires no equilibration, because of
the way it is constructed~\cite{wormhole}.

\section{The end-to-end distance}
\label{sec:eeresults}

\begin{figure}[tp]
  \centering
  \includegraphics[width=\picturewidth]{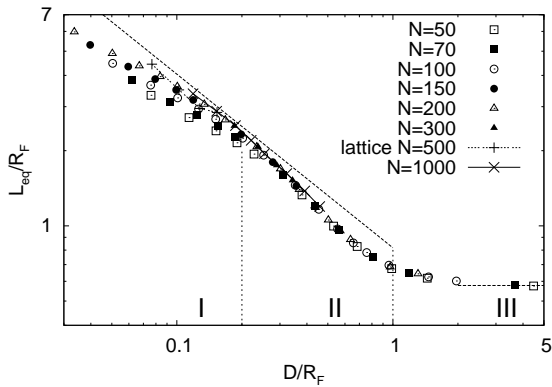}
  \caption{
    Equilibrium chain extension $L_\text{eq}$ (in units of $\reefree$) of a
    cylindrically confined self-avoiding chain as a function of the relative
    confinement $D/\reefree$. The roman numbers denote the three regimes: the
    strong-confinement (I), intermediate (II), and free chain regime (III).  The
    diagonal dashed line in regimes I and II represents the asymptotic power law
    $L_\text{eq} \sim N D^{-0.7}$, the dashed line in regime III corresponds to
    the free chain limit $L_\text{eq}=\sqrt{1/3}\reefree$. The data points for
    $N$ up to 300 are MD results, while those for $N\ge 500$ are MC results. }
  \label{fig:re}
\end{figure}

If a chain is confined in an open cylinder with a diameter much smaller than its
natural size, the chain will be stretched along the cylinder axis.
Fig.~\ref{fig:re} shows the confinement-induced chain stretching
$L_\text{eq}/\reefree$ as a function of the dimensionless parameter
$D/\reefree$. $L_\text{eq}$ denotes the equilibrium end-to-end distance
parallel to the cylinder axis, and $\reefree$ the Flory radius, i.~e. the
unconfined equilibrium end--to--end distance. In the figure, three
different regimes are identified:

\begin{itemize}
\item[I.]  To the region below $D \approx 0.2\reefree$ we refer as the ``strong''
  confinement regime, in which our simulation results are captured reasonably
  well by the scaling prediction $L_\text{eq}\sim ND^{1-1/\nu} \sim ND^{0.7}$.
  The data for large $N$ tend to follow this scaling better, because for small
  chains, the tube diameter approaches the size of the beads, in which case a
  ``blob'' only consists of a few beads.
  
\item[II.] The range between $D\approx 0.2\reefree$ and $D \approx \reefree$,
  we denote as the ``intermediate'' confinement regime. In this regime, we find
  in our simulations a non-negligible chance for the chain to revert its
  direction along the cylinder axis or at least partially fold back one of its
  ends.  Therefore, the average $L$ (or $L_\text{eq}$) decreases faster with
  increasing $D$ than what one may naively expect from the blob picture.
  
\item[III.] In the region above $D \approx \reefree$, the chain is essentially
  free; we call this therefore the ``weak'' confinement regime. Here, the chain
  extension is practically independent of $D$, and is given by
  $L=\sqrt{1/3}\reefree$.
  
\end{itemize}

As we will argue below, the intermediate regime also appears in the behavior of
the relaxation times $\tau$: however, it extends to much smaller values of
$D/\reefree$. Reaching the {\em dynamic} intermediate regime for $\tau$
therefore requires much larger chains than reaching its static counterpart for
$L_\text{eq}$.

Note that, for the same number of monomers, the magnitude of $\reefree$ is
different for the lattice (MC) and off-lattice (MD) models.  The $\reefree$
obtained with the MD simulations is approximately $30\%$ larger than that found
in the MC simulations. This difference can be absorbed into the effective
monomer size $a$, which is proportional to $\reefree$. Because $L_\text{eq}$
scales as $a^{1.7}$, this difference is more pronounced in the estimates of
$L_\text{eq}$: for the same chain length and diameter, the chains in the MD
simulations are about $60\%-100\%$ larger compared to the MC simulations.

%%%%%%%%%%%%%%%%%%%%%%%%%%%%%%%%%%%%%%%%%%%%%%%%%%%%%%%%%%%%%%%%%%%%%%%%%%%%%%
\section{Relaxation time measurements}
\label{sec:rtresults}

\begin{figure}[tp]
  \centering
  \includegraphics[width=\picturewidth]{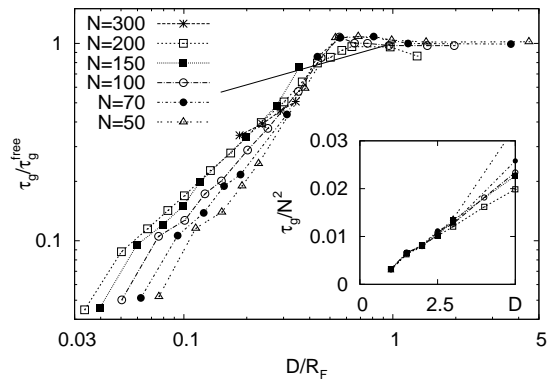}
  \caption{
    Global relaxation time $\tau_\text{g}$ of the radius of gyration $\rg$, in
    units of the relaxation time of a corresponding free chain,
    $\tau_\text{g}^\text{free}$) as a function of the reduced diameter
    $D/\reefree$.  The solid line marks the asymptotic scaling law of $\tau \sim
    N^2D^{0.3}$. The inset shows a graph $\tau_\text{g}/N^2$ vs $D$,
    demonstrating $\tau\sim N^2$ for small diameters.}
  \label{fig:relaxrg}
\end{figure}

We now turn to the global relaxation time $\tau$ of a confined polymer.  In the
case of strong confinement, one expects the blob--scaling result $\tau \sim
N^2D^{2-1/\nu}\approx N^2D^{0.3}$ to be valid, in parallel with the static
scaling law $L_\text{eq} \sim L D^{1-1/\nu}$. For weak confinement or $D \approx
\reefree$, one expects the scaling behavior of the relaxation time to cross over
to that of a free polymer~\cite{DE}: $\tau \sim N^{1+2\nu}\approx N^{2.2}$.
However, little is known about the relaxation time in the {\em dynamic}
intermediate regime. Below we present the relaxation times determined in our
simulations. As it turns out, we are not able to observe the expected
blob--scaling for $\tau$, even for $D\ll 0.3\reefree$. This indicates that the
dynamic intermediate regime spans a much wider parameter space than the static
intermediate regime.

We determine the relaxation time $\tau$ primarily from the slowest exponential
decay of the autocorrelation function
\begin{equation}
\label{eq:CLt}
  C_\text{g}(t)=\frac{\left<  (\rg(t)-\left< \rg \right>)
      (\rg(0)-\left< \rg \right>)\right>}{
    \left<\rg^2\right>-\left<\rg\right>^2},
\end{equation}
where $\rg=(1/N) \sum_i \left(z_i-\left<z_i\right>\right)^2$ denotes the
longitudinal component of the radius of gyration. The slowest exponential decay
$\tau_g$ we obtained from fitting the long-time behavior of $C_\text{g}(t) \sim
e^{-t/\tau_\text{g}}$. $\tau_\text{g}$ turns out to be rather insensitive to the
exact definition of $\rg$; the relaxation time $\tau_L$ of the end-to-end
distance differs from $\tau_\text{g}$ by less than $20\%$, and its overall
behavior is the same ($\tau_L$-data not shown).  We also determined the
relaxation time by measuring the relaxation of a stretched chain.  These
relaxation times agree with the times obtained from the autocorrelation
analysis, as expected on the basis of linear-response theory.  However, the
correlation function of the radius of gyration allows us to obtain better
statistics. The global relaxation time $\tau_\text{g}$ obtained from our MD
simulations is shown in Fig.~\ref{fig:relaxrg}.

For $D>0.5\reefree$, $\tau_\text{g}$ quickly approaches that of a free chain.
However, it has a maximum around $D\approx 0.5\reefree$, beyond which the
relaxation time decays slowly with increasing $D$ and eventually becomes
$D$--independent for sufficiently large $D$. We argue that this non-monotonic
behavior is associated with the reversion of the chain direction or chain
back-folding. This process happens frequently in our simulations for
$D>\reefree$, and its timescale increases with decreasing $D$, as we will show
below. Below $D=0.4\reefree$, however, we do not observe any back-folding in our
trajectories, so that the observed relaxation time is indeed dominated by the
fluctuations of the chain length.

For $D\lessapprox \reefree$, i.e., in the static intermediate and strong
confinement regimes, the relaxation times we find in our MD simulations
approximately scale as
\begin{equation}
\label{eq:tauMD}
\tau_\text{MD} \sim N^{1.75}D^{1.3}.
\end{equation} 
This result has a much stronger $D$--dependence and a slightly weaker
$N$--dependence than the scaling result, which varies as $N^2D^{0.3}$. We refer
to this as a \emph{dynamic} intermediate regime, which spans deeply into the
static strong confinement regime. We will present further analysis of this
behavior based on blob statistics in Sec.~\ref{sec:historesults}.

In a small range in the static intermediate regime, we observe that the
relaxation time increases more steeply with $D$ than indicated by the result in
Eq.~\eqref{eq:tauMD}. However, this range quickly decreases with increasing
chain length, and we will not consider it further. Note that in our simulations,
we cannot consider diameters smaller than $D=2$.  For narrower tubes, two beads
can no longer pass each other without paying an energy penalty. This artifact
significantly reduces the relaxation time, as can be seen from the anomalous
behavior of $\tau_\text{g}$ towards the smallest $D/\reefree$ values in
Fig.~\ref{fig:relaxrg} (see the data points at the left end of each curve).
Also, in this case, $g \approx 1$, and thus the blob picture breaks down.

\begin{figure}[tp]
  \centering
  \includegraphics[width=\picturewidth]{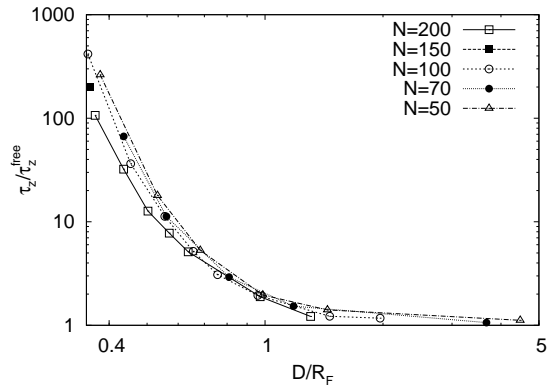}
  \caption{
    Global relaxation time $\tau_z$ obtained from the autocorrelation function
    of $L_z$, i.e., the longitudinal component of the end-to-end vector.  The
    $\tau_z$ is given in units of the relaxation time $\tau_z^\text{free}$ of a
    corresponding free polymer.  Our estimate of $\tau_z$ increases strongly
    with decreasing $D$, in contrast to $\tau_\text{g}$ plotted in
    Fig.~\ref{fig:relaxrg}.}
  \label{fig:relaxsre}
\end{figure}

To illustrate the significance of chain reorientation or back-folding in the weak
confinement regime, we have plotted in Fig.~\ref{fig:relaxsre} our results for
the global relaxation time of the longitudinal end-to-end vector, the difference
$L_z=z_{\text{head}}-z_{\text{tail}}$ of the $z$-coordinates of the head and
tail beads (obviously, $L=|L_z|$). This relaxation time, denoted by $\tau_z$, is
much longer than the $L$- or $\rg$-relaxation times; in our simulations, we were
only able to track it down to $D/\reefree\approx 0.4$.  In the case of a free
polymer, the relaxation times obtained from the fluctuations of its end-to-end
vector and end-to-end distance are expected to be comparable to each other.  In
the presence of confinement, however, the distribution of $L_z$ has two minima
at $L_z \approx \pm L_\text{eq}$, separated by a free energy barrier; these two
minima correspond to the two possible chain orientations. In this case, chain
reorientation, i.~e. ``tunneling'' from one minimum to the other, represents the
slowest mode. It requires a complete back-folding of the chain, which becomes
unlikely with increasing chain length. Since the chain length increases with
decreasing $D$, the reorientation timescale increases fast with $D$.

\begin{figure}[tp]
  \centering
  \includegraphics[width=\picturewidth]{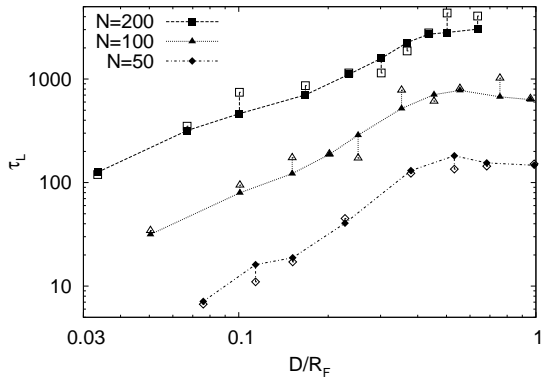}
  \caption{
    Comparison of two relaxation times: the global relaxation time
    $\tau_\text{g}$ of the longitudinal component of the radius of gyration,
    obtained from the autocorrelation function (filled symbols), and that from
    direct stretch-release simulations (open symbols). The dotted lines are
    guides to the eye.}
  \label{fig:stretchrel}
\end{figure}

As an alternative approach, we have determined the relaxation time from direct
stretch--release simulations. In the linear-response regime, the relaxation
times obtained from the autocorrelation function of the chain length should
agree with the results of simulations that measure the relaxation of a slightly
stretched chain. In order to show this, we carried out stretch-release
simulations as follows. We started our simulations with equilibrated chains in
confinement, whose end beads were initially fixed at a distance $L
=L_\text{eq}+2\sigma_L$, where $\sigma_L^2$ is the variance of the chain length
$L$.  We then released the length constraint and recorded the time evolution of
$L$, whose long-time behavior shows an exponential relaxation to its equilibrium
value $L_\text{eq}$: $L(t) - L_\text{eq} \approx e^{-t/\tau_L}$ for large $t$.
In Fig.~\ref{fig:stretchrel}, we have plotted the resulting $\tau_L$, together
with the global relaxation time $\tau_\text{g}$ obtained from the
autocorrelation function.  While the two sets of data agree quite well with each
other, the autocorrelation analysis yields better statistics.  This analysis
demonstrates the reliability of our results for $\tau_\text{g}$.

%%%%%%%%%%%%%%%%%%%%%%%%%%%%%%%%%%%%%%%%%%%%%%%%%%%%%%%%%%%%%%%%%%%%%%%%%%%%%%
\section{Linking between dynamics and statics: the $L$-distribution}
\label{sec:historesults}

\begin{figure}[tp]
  \centering
  \includegraphics[width=\picturewidth]{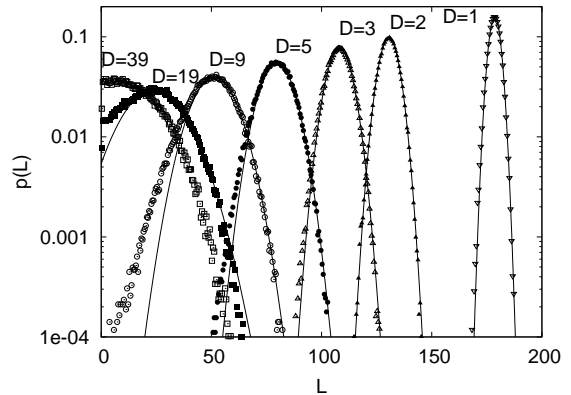}
  \caption{
    MD results for the distribution of the end-to-end distance for $N=200$ and
    various pore sizes. The solid lines are Gaussian fits to the graphs.}
  \label{fig:eehist}
\end{figure}

The Molecular Dynamics simulations, that we have presented so far, are limited
to chain lengths up to $N=300$, because the relaxation of longer chains in
confinement becomes prohibitively slow.  To circumvent this difficulty and to
compliment the MD results, we carried out Monte Carlo (MC) simulations. Note
that one cannot deduce dynamical information directly from the MC wormhole
algorithm, since the artificial dynamics of this method relaxes the system
nonphysically fast.  However, for strong confinement, it is possible to
connect the relaxation time, a dynamic quantity, to a static property, namely
the distribution of the end-to-end distance, as illustrated in
Sec.~\ref{sec:theory}.

In Fig.~\ref{fig:eehist}, we have plotted our MD results for the distribution of
the end-to-end distance $L$, denoted by $p(L)$.  As shown in the figure, this
distribution is essentially Gaussian -- the Gaussian fit works better for small
$D$, i.~e., in the strong confinement regime, as also indicated in
Sec.~\ref{sec:theory}. For large $D$, $p(L)$ is only described well by the
Gaussian in the region on the right side of the peak. The noticeable discrepancy
towards the left end of the distribution can be attributed to partial chain
back-folding, which tends to widen the distribution and creates an exponential
tail.

\begin{figure}[tp]
  \centering
  \includegraphics[width=\picturewidth]{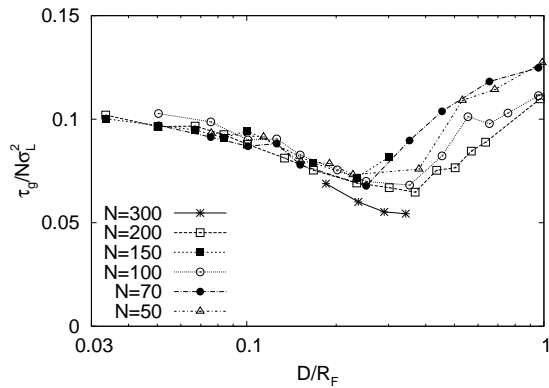}
  \caption{
    Ratio of the relaxation time $\tau_\text{g}$ to $N\sigma_L^2$ obtained from
    MD simulations, as a function of the diameter $D/\reefree$.  As evidenced in
    the figure, this ratio tends to a $D$-independent constant ($\approx 0.1$)
    as $D$ decreases.}
  \label{fig:eefr}
\end{figure}

According to Eq.~\eqref{eq:tauvariance}, the relaxation time is related to
$\sigma^2_L$ via $\tau \sim N \sigma_L^2$ in strong confinement. This means that
in this regime, the $D$ dependence of $\tau_\text{g}$ can be deduced from
$\sigma_L$, since both have the same $D$-scaling. Fig.~\ref{fig:eefr} shows the
ratio of the relaxation time $\tau_\text{g}$ and the variance $N \sigma_L^2$ of
the $L$-distribution, both obtained from the MD simulations. As expected, the
ratio in the figure tends to a $D$-independent constant ($\approx 0.1$) as
$D/\reefree$ decreases. The non-monotonic $D$--dependence around $D\approx
0.3\reefree$ is mainly due to a significant overshoot in $\sigma_L^2$ for
$D\approx 0.5\reefree$. This overshoot is caused by the widening of the
$L$--distribution towards small $L$ due to the onset of chain reorientation or
back-folding. Having established the relation $\tau \sim N \sigma_L^2$, we can
focus on $\sigma_L^2$ and investigate its behavior more closely.

\begin{figure}[tp]
  \centering
  \includegraphics[width=\picturewidth]{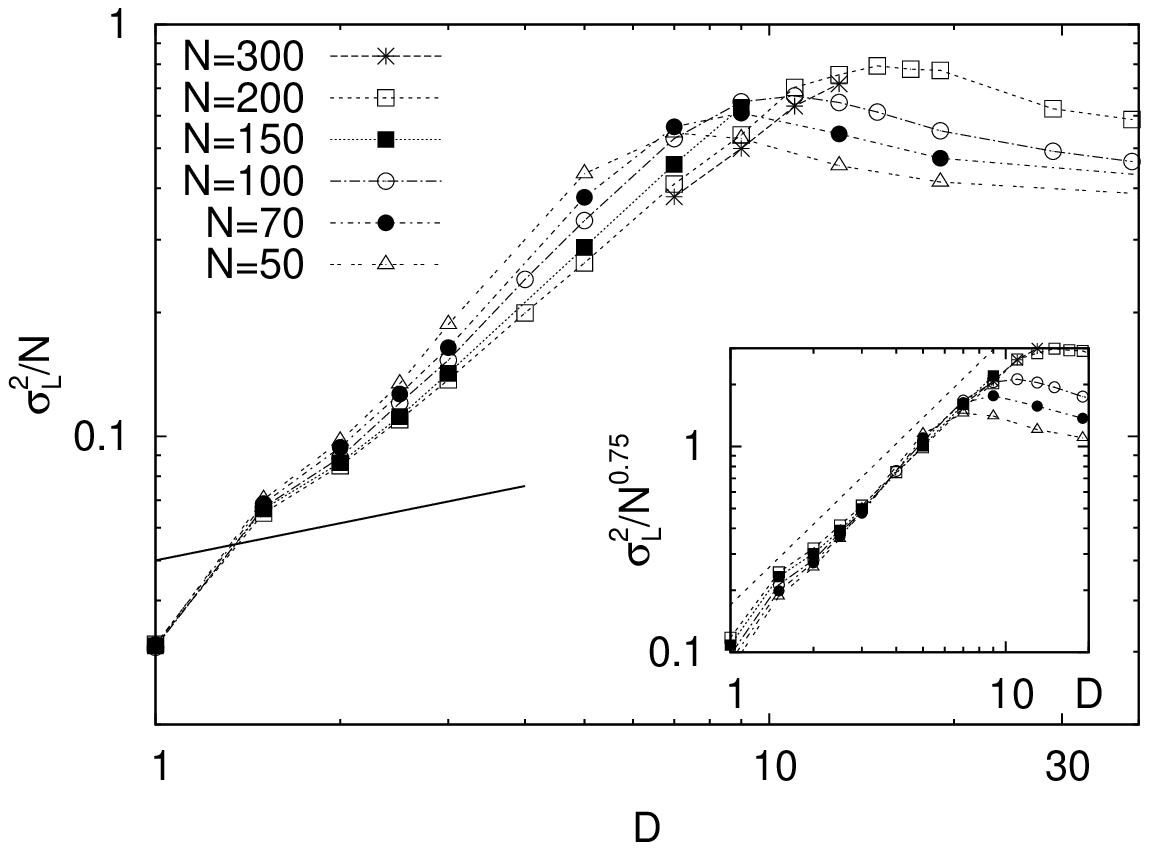}
  \caption{
    MD results for the end fluctuation $\sigma_L^2$ of a confined polymer,
    rescaled by $N$, as a function of $D$. The peaks are all located at
    $D\approx 0.5\reefree$. The solid line represents the $D^{0.3}$-scaling
    expected from the blob approach. In the inset, $\sigma_L^2$ is rescaled by
    $N^{0.75}$, and the dashed line corresponds to $D^{1.3}$, in accordance with
    our best fit.}
  \label{fig:eef}
\end{figure}

\begin{figure}[tp]
  \centering \includegraphics[width=\picturewidth]{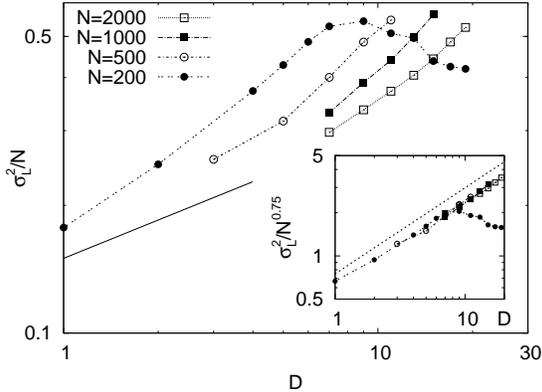}
  \caption{
    MC results for the end fluctuation $\sigma_L^2$ of a confined polymer,
    rescaled by $N$, as a function of $D$. The solid line represents the
    $D^{0.3}$-scaling expected from the blob approach. In the inset,
    $\sigma_L^2$ is rescaled by $N^{0.75}$, and the dashed line corresponds to
    $D^{0.75}$, in accordance with our best fit.}
  \label{fig:eeflat}
\end{figure}

In Fig.~\ref{fig:eef} and \ref{fig:eeflat}, we represent our results for
$\sigma_L^2$ from the MD simulations and the MC lattice simulations,
respectively. The blob picture presented in Sec.~\ref{sec:theory} suggests that
$\sigma_L^2$ scales as $ND^{0.3}$, i.e., $\sigma_L^2$ is extensive.  However,
both simulations indicate that $\sigma_L^2$ is not proportional to $N$, and has
a much stronger $D$-dependence.  More precisely, our data can be fitted well by
$\sigma_L^2\approx N^{0.75} D^{1.3}$ for the MD simulations, and $\sigma_L^2
\approx N^{0.75} D^{0.75}$ for the MC simulations. The main difference between
the two sets of simulations is therefore the $D$--dependence; we will argue
below that this is a lattice/off--lattice effect. While the MD result is
consistent with $\tau$ in Eq.~\ref{eq:tauMD}, the MC result leads to the
estimate
\begin{equation}
  \tau_\text{MC} \sim N^{1.75} D^{0.75}.
\end{equation}
In the next section, we examine this unexpected scaling more carefully. For
this, it proves useful to examine single-blob statistics, especially the
variance $\sigma_b^2$ of the single-blob size distribution. This will enable us
to investigate the $N$-- and $D$--dependence of $\tau$ independently.

\section{Analysis of single blobs}
\label{sec:blobresults}

\begin{figure}[tp]
  \centering
  \includegraphics[width=\picturewidth]{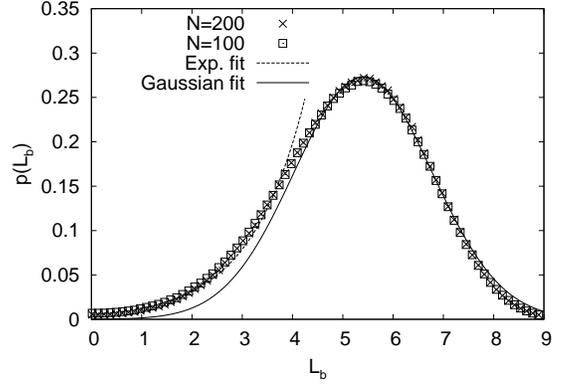}
  \caption{
    Typical blob size distribution for $D=5$ and chain lengths $N=100$ and
    $N=200$. A plain Gaussian distribution (solid line) fits the data well; for
    small $L_b$, the tail is better described by an exponential (dashed line).}
  \label{fig:blobeee}
\end{figure}

To study our system at the level of a single blob, we determine the number of
beads per blob $g$ from our simulations using the relation $L/D=N/g$
(Eq.~\eqref{eq:chainlen}). We then view each sub-chain of beads $i+1, \hdots,
i+g$, consisting of $g$ consecutive beads, as a realization of a blob. For
example, we sample the distance $|z_{i+1}-z_{i+g}|$, i.~e., the longitudinal
distance between two ``end'' beads, as the ``end-to-end'' distance $L_b$ of a
blob. In the strong confinement limit, the average size of a blob constructed
this way is indeed $L_b\approx D$.

Fig.~\ref{fig:blobeee} shows our MD results for a typical blob size distribution
$p(L_b)$, which is well represented by a combination of two functions: a
Gaussian distribution around the peak and an exponential one for small $L_b$.
In this respect, the blob size distribution resembles a short chain in
confinement $\reefree\approx D$, as indicated in Fig.~\ref{fig:eehist}.  In both
cases, the effect of confinement is marginal and chain back-folding is
noticeable, leading to the exponential distribution for small $L_b$.
Furthermore, using both simulations (MD and MC), we found that the distribution
of $L_b$ is practically independent of the chain length if $L>3D$, i.e., if the
chain consists of at least 3 blobs.

\begin{figure}[tp]
  \centering
  \includegraphics[width=\picturewidth]{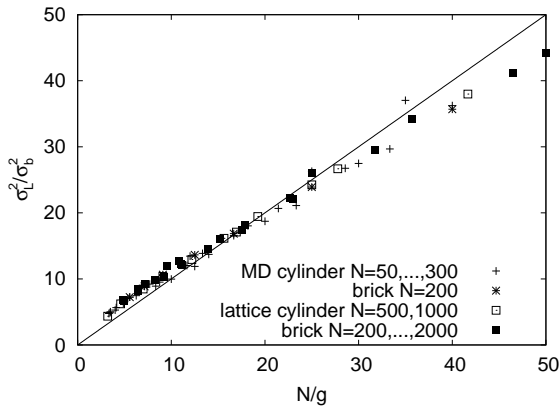}
  \caption{
    MD and MC results for the ratio of the variances of the longitudinal
    end-to-end distance and the blob size, as a function of the number of blobs
    $N_\text{blobs}=N/g=L_\text{eq}/D$. The solid line represents the
    extensiveness $\sigma^2_L=\sigma^2_bN_\text{blobs}$ expected.  Two types
    of confined space were used: a cylinder and a brick, i.~e., a tube with a
    circular and rectangular cross section, respectively.}
  \label{fig:befef}
\end{figure}

\begin{figure}[tp]
  \centering
  \includegraphics[width=\picturewidth]{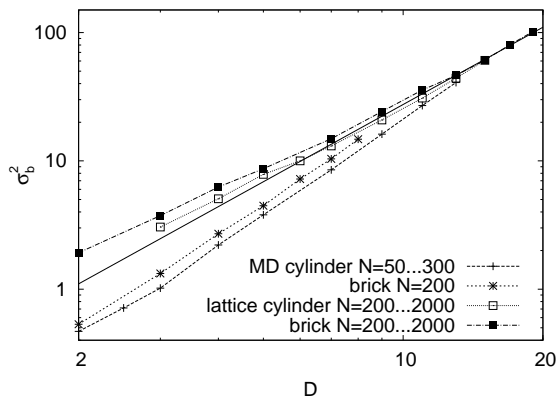}
  \caption{
    MD and MC results for the average variance $\sigma_b^2$ of the single blob
    size, as a function of the tube diameter $D$. As in Fig.~\ref{fig:befef}, we
    used two types of confined space, a cylinder and a brick. The solid line
    represents the theoretically expected $D^2$-scaling.}
  \label{fig:blobeef}
\end{figure}

Fig.~\ref{fig:befef} shows $\sigma_L^2/\sigma_b^2$, the ratio between the
variances of the end-to-end distance and the blob size. If the fluctuations of
the individual blobs are independent, this ratio should equal the number of
blobs $N_\text{blobs} = N/g$, i.~e., the end-to-end fluctuation is extensive.
The data was obtained from both MD and MC simulations, and for two different
shapes of the confined space each: a cylinder and a brick, i.~e., a tube with a
circular and square cross section, respectively. The discrepancy between the
four cases is minor, however, we observe clear deviations from the linear
relationship. The results in Fig.~\ref{fig:befef} for up to $N_\text{blobs}
\approx 30$ are, in fact, better described by $\sigma^2_L/\sigma^2_b\sim
N_\text{blobs}^{0.75}$. For large $N_\text{blobs}$, the data falls onto a line,
however shifted up by about 1 and with a slope of about $0.9$. The shift is a
finite-size effect that simply reflects that the size fluctuations of the two
end--blobs are much larger. The slightly lower slope in turn indicates that the
fluctuations between the blobs are not completely independent. We will discuss
the implications of this finding below.

To further proceed with our discussion regarding the breakdown of extensiveness
of the end fluctuation, we first focus on the variances $\sigma_b^2$, as shown
in Fig.~\ref{fig:blobeef}. We only took into account data points with $L>3D$,
for which the chain length dependence of $L_b$ was negligible. The difference
between the cylinder and the brick is again just an insignificant prefactor.
However, while the shape of the $L_b$-distribution is very similar for the MD
and MC simulations (i.~e., approximately Gaussian), there are significant
differences in the $D$--scaling of $\sigma^2_b$.  The MC lattice simulations
obey the theoretically expected $\sigma_b^2\sim D^2$ law (see
Eqn.~\eqref{eq:keff}) reasonably well, while in the MD simulations, $\sigma_b^2$
rather follows $\sigma_b^2\sim D^{2.6}$.

The most plausible explanation for this difference is through effects near the
cylinder surface. The prediction $\sigma_b^2\sim D^2$ requires that the radial
bead distribution can be written as a function of $\rho/D$ , where $\rho$ is the
radial coordinate~\cite{blob_spring_const}. However, in the off--lattice
simulations, we observe layering effects of the width of about a bead diameter
at the cylinder surface, which are absent on the lattice~\cite{layer}. Because
this surface layer for $D=6$ still contains about 30\% of the total volume, this
makes a significant contribution. Nevertheless, the radial distribution
converges slowly to the universal distribution, and we expect the MD result to
cross over to a $D^2$-law for sufficiently large $D$; unfortunately, this is
computationally inaccessible.

The single-blob analysis presented in this section is consistent with the
unexpected $D$--dependence of $\sigma^2_L$ (and that of the relaxation time): We
observe $\sigma^2_L\approx (N/g)^{0.75}\sigma^2_b=N^{0.75}D^{-1.3}\sigma^2_b$,
where $\sigma^2_b\sim D^{2.6}$ in the MD simulations and $\sigma^2_b\sim D^2$ in
the MC simulations.  In any case, $\sigma_L^2 \propto \sigma_b^2$.  This results
in $\sigma^2_L\approx N^{0.75}D^{1.3}$ for the MD simulation and
$\sigma^2_L\approx N^{0.75}D^{0.7}$ for the MC simulation, which are both in
good agreement with our numerical findings for $\sigma^2_L$. Finally, when
combined with our $\sigma_L$ analysis, the relation \eqref{eq:tauvariance} leads
to a relaxation time $\tau \sim N^{1.75}D^{1.3}$, again in good agreement with
our earlier observations.

\section{Conclusions}
\label{sec:discussion}

We have performed extensive simulations of a self--avoiding polymer under
cylindrical confinement.  In an effort to sample a wide parameter space, we have
complimented our off--lattice MD simulations by lattice MC simulations, and
considered chain lengths $N$ up to $2000$.  Most notably, both types of
simulations have produced global relaxation times $\tau$ that have a much
stronger $D$ dependence than expected from the earlier blob approach~\cite{BdG}.
While our analysis of the longitudinal chain size $L$ for the case $D\lessapprox
0.2\reefree$ tends to support the picture of linear ordering of blobs as assumed
in the blob approach, our results for $\tau$ for the same case $D\lessapprox
0.3\reefree$ do not show the {\em dynamic} blob scaling regime, in which the
blob scaling of $\tau$ holds.
 
Our MD simulation directly measured the time evolution of the end--to--end
distance from which $\tau$ was obtained; in our MC simulation we extracted
$\tau$ from the variance of the end--to--end distance distribution $\sigma_L^2$
through the relation $\tau\sim N\sigma_L^2$, which we have demonstrated to hold
in strong confinement. The absence of the dynamic blob--scaling regime in both
types of simulations has been attributed to finite--size effects.  To this end,
we have examined the chain at the level of a single blob.  In particular, we
have examined the $D$--dependence of $\sigma_b^2$ and examined its relation with
$\sigma_L^2$. Even with up to $N_\text{blobs}=30$ blobs,
$\sigma_L^2=N_\text{blobs}\sigma_b^2$ does not hold. In other words, the
assumption of extensiveness of $\sigma_L$, which underlies the blob picture, is
not satisfied in our simulations. Moreover, the fluctuations of the individual
blobs in our bead--spring model also suffer from finite size effects; in our MD
simulations, we could not reach the expected $\sigma_b^2\sim D^2$ scaling due to
large surface contributions. This indicates that the number of beads per blob
also has to be significantly larger than 100, corresponding to a diameter which
is at least 15 times larger than the effective bead diameter.

Our main findings can be summarized as 
\begin{equation}
  \tau\approx N^{1.75}D^{-1.3}\sigma_b^2.
\end{equation}
When combined with our MD result $\sigma_b^2\approx D^{2.6}$ (obtained for $5\le
N\le 300$ and $2\le D\le 15$), this relation leads to $\tau\approx
N^{1.75}D^{1.3}$. If our MC result $\sigma_b^2 \approx D^2$ (obtained for $200
\le N \le 2000$ and $2\le D\le 20$) is used instead, we obtain $\tau\approx
N^{1.75}D^{0.7}$.  Both are quite different from the asymptotic scaling of
$\tau\sim N^2D^{0.3}$.  Nevertheless, our results for $\tau$ seem to hold for
$0.05 \le D/\reefree \le 1$, and should therefore be experimentally observable.

Importantly, our simulations show that, for reaching the asymptotic scaling
limit with a bead--spring model, about 10,000 beads are necessary.  The aspect
ratio $L/D$, which is identical to the number of blobs, has to be at least 30.
The chain--length requirement for the asymptotic scaling regime $N \gg D^{1/\nu}
\gg 1$ can then be satisfied, however, obtaining the relaxation of such a system
by simulations will remain an intractable task for some time.  For smaller
chains and diameters, the blob-scaling picture has to be used with due caution.
This is similar to recent findings for a polymer stretched by an external
force without confinement~\cite{Thirum}, and for the force--stretching relation
of a polymer in cylindrical confinement~\cite{future2}.

In fact, we notice that even in experiments this regime is hard to reach. For
instance, consider DNA molecules trapped in a cylindrical pore. Although the
Kuhn segments of DNA are cylindrical and therefore our bead--model is not
directly applicable, one can obtain some rough estimate on the minimally
necessary system dimensions. To reach the classical self-avoiding polymer limit
inside a blob of size $D$, this diameter has to be significantly larger than the
persistence length of DNA, which is 50nm. Both our simulations and scaling
analysis~\cite{Rubinstein}, indicate that a pore diameter of about 1$\mu$m is
necessary. The extension of the chain in the pore should be at least 30 times
the pore diameter, which requires a chain length of more than a million base
pairs. This length is significantly larger than the 164,000 base pairs used in a
recent experiment~\cite{reisner}.

%%%%%%%%%%%%%%%%%%%%%%%%%%%%%%%%%%%%%%%%%%%%%%%%%%%%%%%%%%%%%%%%%%%%%%%%%%%%%%
\section*{Acknowledgments}

We thank Sorin Tanase and Kostya Shundyak for many helpful comments and
discussions.  This work is part of the research program of the Stichting voor
Fundamenteel Onderzoek der Materie (FOM), which is supported by the Nederlandse
Organisatie voor Wetenschappelijk Onderzoek (NWO). AA and SJ acknowledge support
from the Marie-Curie program of the European Commission, and SJ the
post-doctoral fellowship from NSERC (Canada).

\appendix
\section{MD simulation parameters}
\label{sec:simparams}

In the following tables, we list the simulation parameters used for the
different MD runs, namely the simulated pore diameters $D$, the equilibration
times $T_{\text{eq}}$ and the sampling times $T_{\text{total}}$, as well as the
time between two successive samples $T_{\text{sample}}$. The number of samples
is hence given by $T_\text{total}/T_\text{sample}$. Times are measured in
multiples of the microscopic time $\tau_{WCA}$, which corresponds to 100 MD time
steps, i.~e. the total number of performed time steps is equal to
$100T_\text{total}$.

\newlength\entrywidth
\newlength\topsepa
\setlength\topsepa{0.1em}
\newlength\presepa
\setlength\presepa{0.5em}
\settowidth\entrywidth{0.00}
\addtolength\entrywidth{0.25em}

\raggedright
\vspace{\presepa}
\noindent N=50\\[\topsepa]
  \begin{tabular}{|l%
      |p{\entrywidth}|p{\entrywidth}|p{\entrywidth}|p{\entrywidth}%
      |p{\entrywidth}|p{\entrywidth}|p{\entrywidth}|p{\entrywidth}|}
    \hline
    $D$
    & 2 & 2.5 & 3 & 3.5 %
    & 4 & 6 & 8 & 10 \\
    \hline
    $T_{\text{eq}}/10^3$
    & 1.2 & 4 & 4 & 4 %
    & 3.2 & 5.2 & 8 & 10.8 \\
    \hline
    $T_{\text{total}}/10^6$
    & 0.30 & 1.00 & 1.00 & 1.00 %
    & 0.80 & 3.91 & 6.01 & 8.11 \\
    \hline
    $T_{\text{sample}}$
    & 3 & 10 & 10 & 10 %
    & 8 & 13 & 20 & 27 \\
    \hline
  \end{tabular}
  \vspace{\topsepa}\\
  \begin{tabular}[b]{|l%
      |p{\entrywidth}|p{\entrywidth}|p{\entrywidth}|}
    \hline
    $D$
    & 14 & 20 & 60 \\
    \hline
    $T_{\text{eq}}/10^3$
    & 9.95 & 9.95 & 20 \\
    \hline
    $T_{\text{total}}/10^6$
    & 10.8 & 10.8 & 10.0 \\
    \hline
    $T_{\text{sample}}$
    & 50 & 50 & 50 \\
    \hline
\end{tabular}

\vspace{\presepa}
\noindent N=70\\[\topsepa]
  \begin{tabular}[b]{|l%
      |p{\entrywidth}|p{\entrywidth}|p{\entrywidth}|p{\entrywidth}%
      |p{\entrywidth}|p{\entrywidth}|p{\entrywidth}|p{\entrywidth}|}
    \hline
    $D$
    & 2 & 2.5 & 3 & 3.5 %
    & 4 & 6 & 8 & 10 \\
    \hline
    $T_{\text{eq}}/10^3$
    & 2.4 & 3.2 & 4 & 5.2 %
    & 6 & 10.4 & 16 & 20.8 \\
    \hline
    $T_{\text{total}}/10^6$
    & 0.60 & 0.80 & 1.00 & 1.31 %
    & 4.51 & 7.81 & 12.0 & 15.6 \\
    \hline
    $T_{\text{sample}}$
    & 6 & 8 & 10 & 13 %
    & 15 & 26 & 40 & 52 \\
    \hline
  \end{tabular}
  \vspace{\topsepa}\\
  \begin{tabular}[b]{|l%
      |p{\entrywidth}|p{\entrywidth}|p{\entrywidth}|}
    \hline
    $D$
    & 14 & 20 & 60 \\
    \hline
    $T_{\text{eq}}/10^3$
    & 20.8 & 100 & 40 \\
    \hline
    $T_{\text{total}}/10^6$
    & 15.6 & 15.6 & 20.0 \\
    \hline
    $T_{\text{sample}}$
    & 52 & 52 & 100 \\
    \hline
  \end{tabular}

\vspace{\presepa}
\noindent N=100\\[\topsepa]
  \begin{tabular}[b]{|l%
      |p{\entrywidth}|p{\entrywidth}|p{\entrywidth}|p{\entrywidth}%
      |p{\entrywidth}|p{\entrywidth}|p{\entrywidth}|p{\entrywidth}|}
    \hline
    $D$
    & 2 & 2.5 & 3 & 3.5 %
    & 4 & 5 & 6 & 8 \\
    \hline
    $T_{\text{eq}}/10^3$
    & 4.8 & 6.8 & 8.8 & 10.8 %
    & 12.4 & 16 & 22 & 28 \\
    \hline
    $T_{\text{total}}/10^6$
    & 1.20 & 1.71 & 2.21 & 2.71 %
    & 9.31 & 12.0 & 16.5 & 21.0 \\
    \hline
    $T_{\text{sample}}$
    & 12 & 17 & 22 & 27 %
    & 31 & 40 & 55 & 70 \\
    \hline
  \end{tabular}
  \vspace{\topsepa}\\
  \begin{tabular}[b]{|l%
      |p{\entrywidth}|p{\entrywidth}|p{\entrywidth}|p{\entrywidth}%
      |p{\entrywidth}|p{\entrywidth}|p{\entrywidth}|}
    \hline
    $D$
    & 10 & 12 & 14 & 16 %
    & 20 & 30 & 40 \\
    \hline
    $T_{\text{eq}}/10^3$
    & 34 & 100 & 300 & 500 %
    & 500 & 800 & 1000 \\
    \hline
    $T_{\text{total}}/10^6$
    & 25.5 & 30.0 & 30.0 & 30.0 %
    & 30.0 & 29.4 & 30.0 \\
    \hline
    $T_{\text{sample}}$
    & 85 & 100 & 100 & 100 %
    & 100 & 100 & 100 \\
    \hline
  \end{tabular}

\vspace{\presepa}
\noindent N=150\\[\topsepa]
  \begin{tabular}[b]{|l%
      |p{\entrywidth}|p{\entrywidth}|p{\entrywidth}|p{\entrywidth}%
      |p{\entrywidth}|p{\entrywidth}|p{\entrywidth}|p{\entrywidth}|}
    \hline
    $D$
    & 2 & 2.5 & 3 & 3.5 %
    & 4 & 6 & 8 & 10 \\
    \hline
    $T_{\text{eq}}/10^3$
    & 11.2 & 15.2 & 19.2 & 24 %
    & 28 & 48 & 72 & 80 \\
    \hline
    $T_{\text{total}}/10^6$
    & 2.81 & 3.82 & 4.82 & 6.02 %
    & 7.03 & 12.0 & 7.98 & 7.57 \\
    \hline
    $T_{\text{sample}}$
    & 28 & 38 & 48 & 60 %
    & 70 & 120 & 180 & 200 \\
    \hline
  \end{tabular}

\vspace{\presepa}
\noindent N=200\\[\topsepa]
  \begin{tabular}[b]{|l%
      |p{\entrywidth}|p{\entrywidth}|p{\entrywidth}|p{\entrywidth}%
      |p{\entrywidth}|p{\entrywidth}|p{\entrywidth}|p{\entrywidth}|}
    \hline
    $D$
    & 2 & 2.5 & 3 & 3.5 %
    & 4 & 5 & 6 & 8 \\
    \hline
    $T_{\text{eq}}/10^3$
    & 50 & 50 & 50 & 50 %
    & 50 & 50 & 50 & 56 \\
    \hline
    $T_{\text{total}}/10^6$
    & 2.89 & 4.09 & 5.30 & 5.42 %
    & 6.22 & 8.03 & 9.94 & 13.4 \\
    \hline
    $T_{\text{sample}}$
    & 24 & 34 & 44 & 54 %
    & 62 & 80 & 110 & 140 \\
    \hline
  \end{tabular}
  \vspace{\topsepa}\\
  \begin{tabular}[b]{|l%
      |p{\entrywidth}|p{\entrywidth}|p{\entrywidth}|p{\entrywidth}%
      |p{\entrywidth}|p{\entrywidth}|p{\entrywidth}|p{\entrywidth}|}
    \hline
    $D$
    & 10 & 12 & 14 & 16 %
    & 18 & 20 & 30 & 40 \\
    \hline
    $T_{\text{eq}}/10^3$
    & 56 & 56 & 56 & 60 %
    & 60 & 100 & 100 & 300 \\
    \hline
    $T_{\text{total}}/10^6$
    & 14.5 & 14.1 & 14.1 & 15.1 %
    & 15.1 & 12.1 & 12.1 & 12.1 \\
    \hline
    $T_{\text{sample}}$
    & 170 & 140 & 140 & 150 %
    & 150 & 200 & 200 & 200 \\
    \hline
  \end{tabular}

\vspace{\presepa}
\noindent N=300\\[\topsepa]
  \begin{tabular}[b]{|l%
      |p{\entrywidth}|p{\entrywidth}|p{\entrywidth}|p{\entrywidth}%
      |p{\entrywidth}|p{\entrywidth}|p{\entrywidth}|p{\entrywidth}|}
    \hline
    $D$
    & 8 & 10 & 12 & 14 \\
    \hline
    $T_{\text{eq}}/10^3$
    & 100 & 100 & 1000 & 2000 \\
    \hline
    $T_{\text{total}}/10^6$
    & 30.1 & 30.1 & 30.1 & 30.1 \\
    \hline
    $T_{\text{sample}}$
    & 200 & 200 & 200 & 200 \\
    \hline
  \end{tabular}

\vspace{\presepa}
\noindent N=200, box shaped simulation box\\[\topsepa]
  \begin{tabular}[b]{|l%
      |p{\entrywidth}|p{\entrywidth}|p{\entrywidth}|p{\entrywidth}%
      |p{\entrywidth}|p{\entrywidth}|p{\entrywidth}|p{\entrywidth}|}
    \hline
    $D$
    & 2 & 3 & 4 & 5 %
    & 6 & 7 & 8 & 9 \\
    \hline
    $T_{\text{eq}}/10^3$
    & 9.6 & 17.6 & 24.8 & 32 %
    & 40 & 44 & 48 & 52 \\
    \hline
    $T_{\text{total}}/10^6$
    & 2.41 & 4.42 & 6.22 & 8.03 %
    & 10.0 & 11.0 & 12.0 & 13.1 \\
    \hline
    $T_{\text{sample}}$
    & 24 & 44 & 62 & 80 %
    & 100 & 110 & 120 & 130 \\
    \hline
  \end{tabular}
  \vspace{\topsepa}\\

  \begin{tabular}[b]{|l%
      |p{\entrywidth}|p{\entrywidth}|p{\entrywidth}|p{\entrywidth}|}
    \hline
    $D$
    & 10 & 12 & 14 & 16 \\
    \hline
    $T_{\text{eq}}/10^3$
    & 52 & 56 & 56 & 60 \\
    \hline
    $T_{\text{total}}/10^6$
    & 13.1 & 14.1 & 14.1 & 15.1 \\
    \hline
    $T_{\text{sample}}$
    & 130 & 140 & 140 & 150 \\
    \hline
  \end{tabular}

%%%%%%%%%%%%%%%%%%%%%%%%%%
%%%%%%%%%%%% Literature
%%%%%%%%%%%%%%%%%%%%%%%%%%
%\bibliographystyle{aip}


\begin{thebibliography}{20}
\bibitem{footflory} The original Flory exponent derived at the meanfield
  level~\cite{Flory} is $\nu =3/5$, but we use the more accurate estimate~\cite{DE} $\nu =
  0.59$.
\bibitem{blob_spring_const} The elastic free energy per blob, $\mathcal{F}_b$,
  can be Taylor--expanded for small deformations in powers of the longitudinal blob
  deformation $\Delta L_b$. One obtains~\cite{BdG} $\Delta\mathcal{F}_b/k_BT \sim
  (\Delta L_b/D)^2 + \text{higher-order
    terms}$, which leads to the blob spring constant given in the text.
\bibitem{footexp} If $\nu=3/5$ were used instead as in Ref~\cite{BdG}, one would
  reproduce the scaling result $\tau \sim N^2 D^{1/3}$ instead.
\bibitem{layer} This layering is a consequence of the different shapes of the
  bead--bead radial distribution function. On the lattice, it is nonphysically
  flat, but off--lattice, it shows a significant peak at contact.
\end{thebibliography}
\end{document}